\documentclass[prl,twocolumn,showpacs,preprintnumbers,amsmath,amssymb]{revtex4}

\usepackage{graphicx}% Include figure files
\usepackage{dcolumn}% Align table columns on decimal point
\usepackage{bm}% bold math
\usepackage{amsmath}
\usepackage{latexsym}

\newcommand{\be}{\begin{equation}}
\newcommand{\ee}{\end{equation}}

\newcommand{\req}[1]{Eq.~(\ref{#1})}

\newcommand{\rfig}[1]{Fig.~\ref{#1}}
\newcommand{\rref}[1]{Ref.~\cite{#1}}

\begin{document}

\title{Weak anti-localization in epitaxial graphene: \\ evidence for chiral electrons}

\author{Xiaosong Wu$^1$, Xuebin Li$^1$, Zhimin Song$^1$, Claire Berger$^{1,2}$, Walt A. de Heer$^{1}$}
\affiliation{
$^1$School of Physics, Georgia Institute of Technology, Atlanta, GA 30332 \\
$^2$CNRS - Institut N\'{e}el, BP 166, 38042 Grenoble cedex 9, France}

\date{\today}

\begin{abstract}

Transport in ultrathin graphite grown on silicon carbide is dominated by the electron-doped epitaxial layer at the interface. Weak anti-localization in 2D samples manifests itself as a broad cusp-like depression in the longitudinal resistance for magnetic fields 10 mT$< B <$ 5 T. An extremely sharp weak-localization resistance peak at $B=0$ is also observed. These features quantitatively agree with graphene weak-(anti)localization theory implying the chiral electronic character of the samples. Scattering contributions from the trapped charges in the substrate and from trigonal warping due to the graphite layer on top are tentatively identified. The Shubnikov-de Haas oscillations are remarkably small and show an anomalous Berry's phase.
\end{abstract}


\maketitle

The extraordinary transport properties of carbon nanotubes \cite{Frank1998}, graphene \cite{Berger2004,Novoselov2005,Zhang2006} and graphene ribbons \cite{Berger2006} make this an exciting and promising new material for nanoelectronics. Epitaxial graphene (EG) grown on silicon carbide substrates can be patterned using standard lithography methods and is robust and reproducible. Their transport properties reflect those previously observed in deposited microscopic single graphene sheets from exfoliated graphite. In both cases, the special transport properties have their origin in the chiral nature of the charge carriers, causing reduced backscattering \cite{Ando1998} and an anomalous quantum Hall effect \cite{Gusynin2005,Peres2006}. Chirality is due to the equivalence of the A and B sublattices of graphene, which produces two inequivalent Dirac cones at opposite corners of the Brillouin zone at $\mathbf{K}$ and $\mathbf{K'}$. Consequently, the wave functions have an additional isospin quantum label. On a specific cone the isospins of oppositely directed electrons are also reversed.  Hence, isospin conserving (IC) scattering processes ({\it i.e.} long-range scatterers that do not distinguish between A and B atoms) cannot backscatter the charge carriers, reducing the resistance caused by those scatterers. The effect diminishes in a magnetic field, resulting in a positive magnetoresistance. This weak anti-localization (WAL) effect is a signature of the isospin. It stands in contrast to the usual the weak localization (WL) effect, characterized by a negative magnetoresistance, which occurs in these materials at point defects that locally break the sublattice degeneracy, thereby causing intervalley scattering (from one Dirac cone to the other).  Note that due to the A-B stacking in graphite, the sublattice degeneracy is lifted so that WAL is neither expected nor observed.

We show here evidence for WAL in two-dimensional thin graphitic layers grown on single crystal silicon carbide. The effect is quantitatively in agreement with recent graphene WAL theory demonstrating the isospin character of the charge carriers in epitaxial graphene \cite{McCann2006}. In patterned, quasi one-dimensional epitaxial graphene structures we recently demonstrated quantum confinement effects, exceptionally high mobilities and evidence for an anomalous Berry's phase \cite{Berger2006}.  Transport is dominated by the interface layer, which due to the built-in electric field, is charged with an electron density $n=3.4\times10^{12}$ cm$^{-2}$.

In two-dimensional graphene, depending on the relative magnitude of the intervalley scattering time $\tau_{iv}$ and the phase coherence time $\tau_\phi$, either WL or WAL has been predicted \cite{Ando1998}. The phase interference correction to the resistance depends on the nature of the disorder \cite{McCann2006,Morpurgo2006}. In \rref{McCann2006}, McCann {\it et al.} point out that WAL is suppressed by any scattering mechanism that changes isospin, as well as the warping term in the Hamiltonian.

For epitaxial graphene samples in this study, elastic scattering favorable for WAL can be caused by remote charges like the counterions in the substrate. On the other hand, atomically sharp disorder ({\it i.e.} local defects and edges) causes intervalley scattering, and gives rise to WL. Finally, trigonal warping tends to suppress the WAL. Trigonal warping can be caused by the interlayer interactions of a graphene sheet on top of the EG layer. Hence scattering from each of the three components of the EG system is expected to contribute to the MR.

In the case that IC scattering dominates, the correction to the sheet magnetoresistance can be expressed as \cite{McCann2006}:
\be \label{wal}
\begin{split}
\Delta \rho(B)& =-\frac{e^2\rho^2}{\pi h}\left[ F\left(\frac{2\tau_\phi}{\tau_B}\right)-F\left(\frac{2}{\tau_B(\tau_\phi^{-1}+2\tau_{iv}^{-1})}\right) \right. \\
& \left. \qquad\qquad -2F\left(\frac{2}{\tau_B(\tau_\phi^{-1}+\tau_{iv}^{-1}+\tau_w^{-1})}\right) \right]
\end{split}
\ee
\begin{displaymath}
F(z)=\ln z+\Psi\left(\frac{1}{2}+\frac{1}{z}\right).
\end{displaymath}

In contrast, the correction to the magnetoresistance in conventional 2D metals due to the weak-localization is given by \cite{Beenakker1991}:
\be \label{wl}
\Delta \rho(B)=-\frac{2e^2\rho^2}{\pi h}\left[ F\left (\frac{2\tau_\phi}{\tau_B} \right)- F\left(\frac{2\tau}{\tau_B} \right) \right].
\ee
Here $\Psi$ is the digamma function, $\tau$ is the transport time \cite{McCann2006}, $\tau_w$ is the warping-induced relaxation time, $\tau_B=\hbar/2eDB$, where $D$ is the diffusion constant. \req{wal} shows that the amplitude of the WL peak at $B=0$ depends on $\tau_{iv}$. Another important implication is that WAL can manifest itself in relatively high magnetic fields ({\it i.e.} where mainly short return trajectories contribute to the interference correction), even in the presence of significant intervalley scattering. Qualitatively, in that case the long tail of the negative MR will ultimately give way to a positive MR at high fields. In other words, the MR will have a sharp peak at $B=0$ and a positive slope at higher fields. All of these interference corrections to the resistance diminish with increasing temperature due to the reduction of the phase coherence time.

An earlier study of a quasi-1D EG ribbon shows a high mobility and a long coherence length \cite{Berger2006}. In that case, edge scattering was shown to dominate the electron transport. Also a sharp MR peak near $B=0$, consistent with 1D WL, was observed. This is consistent with intervalley scattering from the graphene edges which is favorable for producing a WL peak.

In this Letter we report evidence for WAL in 2D EG. Due to the epitaxial growth, the EG layer is of high crystalline quality \cite{Haas2006} and protected from the environment by the graphite over layer. Ultra-thin graphite layers were made by thermal decomposition of single-crystal silicon carbide on the 000$\overline{1}$ face. Unlike graphite, which consists of A-B stacked graphene planes, our samples show evidence for azimuthal orientational disorder as seen in low energy electron diffraction and grazing incidence X-ray diffraction \cite{stacking}. The graphene symmetry is preserved. Graphite layers were patterned to produce a standard Hall bar 100 $\mu$m$\times$1000 $\mu$m in size. The wired samples were placed into a He$^4$ cryostat providing temperatures down to 1.4 K. Standard four-point measurements were carried out using a lock-in amplifier. A magnetic field was applied perpendicular to the graphene layer. Applying a magnetic field parallel to the graphene layer diminished the magnetoresistance to 1.5 \% of its perpendicular-field value. This establishes an upper limit to the angular dispersion of the normal EG layer of $\delta\theta \leq 0.015$. A detailed analysis on the WL peak in a perpendicular and parallel magnetic fields gives a smaller upper limit : $\delta\theta \leq 0.006$. The transport electron density 3.8$\times$10$^{12}$ cm$^{-2}$ was determined from the Shubnikov-de Haas (SdH) oscillations. The low field Hall slope is 137 $\Omega$/T, corresponding to a Hall electron density of $n_H=4.6\times$10$^{12}$ cm$^{-2}$, which agrees with the density from SdH oscillations.

In contrast to 1D samples, in 2D films, edge scattering is significantly reduced and universal conductance fluctuations are absent. The WL is strongly suppressed and a positive MR in the intermediate field region is observed. These properties are reflected in the model developed by McCann {\it et al.} (\req{wal}).

\begin{figure}
\includegraphics[width=0.36\textwidth]{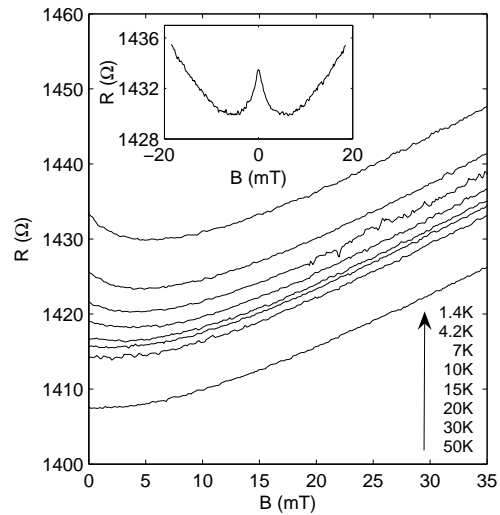}
\caption{\label{MR} Low-field magnetoresistance for a sample of 100 $\mu$m by 1000 $\mu$m at various temperatures. The inset: the magnetoresistance peak near $B=0$ can be seen when sweeping the field from -20 mT to 20 mT at 1.4K. The electron mobility is 11600 cm$^2$/V~s. The transport time $\tau\sim0.26$ ps is inferred from the resistivity. }
\end{figure}

We have studied two samples and both displayed similar behavior. \rfig{MR} shows the low-field resistance ($B<35$ mT) of one sample for temperatures ranging from 1.4 K to 50 K, which consists of a temperature dependent peak around zero field and a temperature independent parabolic component. As conventional, to bring out the temperature dependent part of the MR, the 50 K data (for which the peak is absent) is subtracted from the lower temperature data $R^*(B,T)=R(B,T)-R(B,50K)$ and then the temperature dependent MR is calculated by $\Delta R^*=R^*(B,T)-R^*(0,T)$. An extremely sharp WL peak at $B=0$ (upper-right inset of \rfig{fit}) indicates a long phase coherence length (see below). As expected, the amplitude of the WL peak decreases while its width increases with increasing temperature as a result of the reduction of the phase coherence length. An attempt to fit the WL peak using conventional WL theory is shown in \rfig{fit}. Here a scaling factor $\alpha\sim0.3$ was introduced to obtain a reasonable fit about $B=0$. This factor suggests that the WL is suppressed \cite{Morozov2006}.

\begin{figure}
\includegraphics[width=0.4\textwidth]{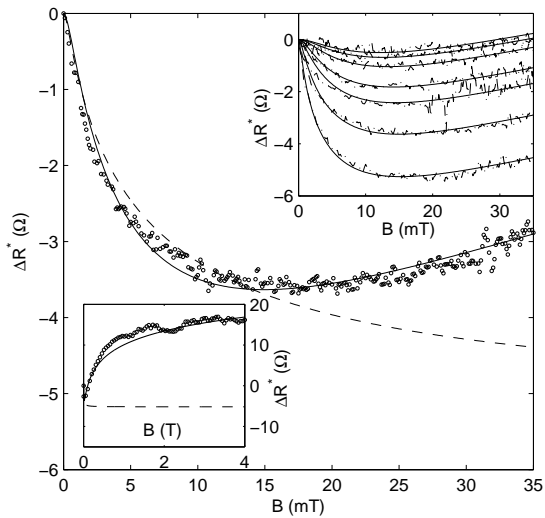}
\caption{\label{fit} Fit of low-field $\Delta R^*$ at 4.2 K to two models. $\Delta R^*$ is the magnetoresistance after subtracting a background as described in the main text. Open circles are the data. Dashed line is a fit to the WL theory for a normal 2D metal (\req{wl}). A scaling factor $\alpha\sim0.3$ is introduced in order to fit data, nevertheless an acceptable fit can only be obtained below 20 mT. Solid line is a fit to the model by McCann {\it et al.} (\req{wal}). The fit shows a good agreement for the entire range of field. Right inset: dash-dot, $\Delta R^*$ for $T=$1.4 K, 4.2 K, 7 K, 10 K, 15 K, 20 K, 30 K from bottom to top, solid line, fits to \req{wal}. Left inset: plot of the main panel extended to 4 T.}
\end{figure}

\begin{figure}
\includegraphics[width=0.3\textwidth]{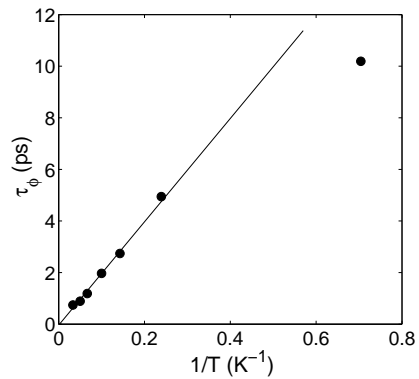}
\caption{\label{TauT} Phase coherence time $\tau_\phi$ as a function of inverse temperature $1/T$. $\tau_\phi$ is obtained by a best fit to \req{wal} for each temperature while $\tau_{iv}$ and $\tau_w$ are the same for all temperatures. Solid line is a guide to the eye.}
\end{figure}

However an important deviation in $\Delta R^*$ from conventional WL theory is observed at fields above 20 mT where $\Delta R^*$ increases with increasing field up to $B=4$ T. This temperature dependent positive MR is a clear signature of WAL as discussed above \cite{eeinteraction}. Moreover, the entire MR behavior is described very well by \req{wal}. Both the high field behavior of the MR, which is dominated by WAL for $B>20$ mT, and the low field behavior, for $B<20$ mT which is dominated by WL \cite{McCann2006}, are convincingly reproduced (the fits are shown in \rfig{fit} and the left inset). To produce this fit we reasonably assumed that $\tau_{iv}$ and $\tau_w$ are temperature independent, while $\tau_\phi$ is temperature dependent. We find that $\tau_{iv}$ and $\tau_w$ are 1.0 ps and 0.28 ps, respectively. The temperature dependence of the phase coherence time $\tau_\phi$  is plotted in \rfig{TauT}. The coherence time is proportional to the inverse temperature, which, in 2D, is an indication that the dominant phase-breaking mechanism is $e$-$e$ scattering. This is consistent with our previous results \cite{Berger2006}and provides additional support for \req{wal}. (The deviation from linearity at the lowest temperature suggests a saturation of the coherence time. This is seen not only in our samples \cite{Berger2006} but in 2D electron gases in general \cite{Lin2002}. There is still no consensus on its origin). To the best of our knowledge this constitutes the first experimental evidence for WAL in graphene \cite{corrugations}. We reiterate that graphite shows normal WL \cite{Morozov2006,QIEgraphite} so that the WAL relates to the epitaxial graphene layer, and consequently these layers can be used to study the intriguing properties of chiral electrons.

The temperature dependence of the $e$-$e$ scattering time can be expressed as \cite{Abrahams1981}:
\be
\tau_{ee}=-\frac{2E_F\tau}{k_B\ln(T/T_1)}\frac{1}{T}
\label{tauee}
\ee
where $E_F$ is the Fermi energy, $k_BT_1=\hbar^3D^3\kappa^4\epsilon^2/e^4$, $\epsilon$ and $\kappa$ are the dielectric constant and screening constant, respectively. At low temperatures, the $\ln T$ dependence is usually negligible and $\tau_{ee}\simeq C/T$. The screening constant in 2D is given by $\kappa=8\pi Ne^2/\epsilon$, where $N$ is the density of states per spin per valley. For graphene, $N=k_F/hv_F$ and $\epsilon\approx 3.28$. So, we estimated $\kappa\approx9.2\times 10^6$ cm$^{-1}$. We find that $T_1\approx 2.7\times 10^{10}$ K so that $C\approx63$ ps~K, which is of the same order as the value of 20 ps~K obtained from the plot in \rfig{TauT}. Considering the complexity of \req{tauee}, this is a remarkable agreement, which, again, supports the fit to \req{wal}.

Apparently IC scattering is the dominant elastic scattering mechanism in 2D epitaxial graphene giving rise to the extended WAL depression. It is likely to be caused by the long-range interactions with the counter-ions (``holes") in the substrate. The isospin symmetry breaking, intervalley scattering events are apparently relatively rare ({\it i.e.} scattering times are long), since WL effects are only observed at weak fields indicating that only long return trajectories involve such a scattering events. This is consistent with the high crystalline purity and the absence of edges in the 2D material. In contrast, in 1D samples edge scattering dominates. On the other hand, it should be expected that the IC scattering becomes ineffective in quasi-1D ribbons due to forbidden back-scattering \cite{Ando1998}, (which is why carbon nanotubes tend to be ballistic conductors \cite{Frank1998}). In fact we were not able to detect WAL in 1D ribbons ({\it e.g.} \rref{Berger2006}).

\begin{figure}
\includegraphics[width=0.4\textwidth]{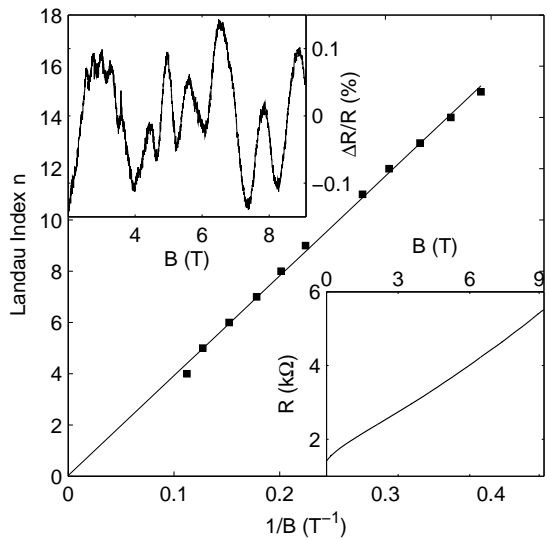}
\caption{\label{SdH} Landau plot (Landau index $n$ vs $1/B$) of the peaks of the SdH oscillations up to 9 T at 4.2 K (solid square). The solid line corresponds to a linear fit, which gives a magnetic frequency of 39 T and an intercept of $0.03\pm0.17$ on the $n$ axis, indicating the anomalous Berry's phase. Right inset: resistance as a function of the magnetic field at $T=4.2$ K. Left inset: small SdH oscillations revealed after subtracting a quadratic background from the measured resistance that is plotted in the right inset.}
\end{figure}

After subtracting a quadratic background, extremely small but well-defined SdH oscillations (less than 0.2\%) are observed in the high field resistance at 4.2 K (insets of \rfig{SdH}). The Landau plot for the maxima of the resistance SdH oscillations (\rfig{SdH}) shows a linear trend yielding a magnetic frequency $B_0=39$ T, from which we have deduced the carrier density. The almost zero intercept on the $n$ axis indicates a Berry phase of $\pi$, consistent with graphene (see also \rref{Berger2006}). In this case, the small amplitude of SdH oscillations cannot be caused by a short scattering time $\tau$ because $\tau\sim0.26$ ps, is longer than those in \rref{Novoselov2005}, where SdH oscillations even evolved into a quantum Hall effect. More striking is that the SdH amplitudes in 2D samples are much smaller than the SdH amplitudes in 1D samples ({\it c.f.} \rref{Berger2006}). We find that the amplitudes are approximately inversely proportional to the sample width. This observation leads to the interesting speculation that in graphene scattering from atomically sharp defects enhances the SdH oscillation amplitudes, and consequently, that such local defects are also required to produce the integer quantum Hall effect. Apparently in our 2D samples the defect density of intervalley scatterers in the bulk is too low to produce large SdH oscillations or an integer quantum Hall effect.

In summary, epitaxial graphene exhibits a number of graphene properties. The isospin degeneracy that causes both weak anti-localization as well as the anomalous Berry's phase is a graphene property that is not present in graphite. It appears that the 2D bulk scattering in EG causes WAL and is dominated by valley symmetry conserving processes, consistent with scattering from long range potentials arising from charges in the substrate. WL appears to be primarily caused by isospin symmetry breaking scattering, for example from edges. Trigonal warping scattering due to the graphite layer on top of the EG could explain the reduction of the WAL. The extremely small SdH amplitudes (which preclude the quantum Hall effect) may be directly related to the scattering processes in the 2D bulk EG.

Supported by NSF-NIRT grant 0404084, NSF-MRI grant 0521041, a grant from Intel, and a USA-France travel grant from CNRS.


\begin{thebibliography}{99}
\bibitem{eeinteraction} Electron-electron interaction can also lead to a temperature dependent positive MR, provided the magnetic field is large enough so that either $g\mu_BB > k_BT$ or $\hbar D/l_{cy}^2 > k_BT$ is satisfied, here $l_{cy}$ is the cyclotron radius. The positive MR observed in our experiment starts from 20 mT. To see any effect of $e$-$e$ interaction at this field, we have to go to a temperature below 30 mK, according to the above condition. Consequently, the effect of $e$-$e$ interaction can be ruled out.
\end{thebibliography}
\end{document}